\documentclass[preprint,12pt]{elsarticle}
\usepackage{graphicx}
\usepackage{amssymb}
\usepackage{amsmath}

\journal{Physics Letters B} 

\def\lamb#1#2{$^{#1}_{\Lambda}${#2}}
\def\lam#1#2{$^{#1}_{~\Lambda}${#2}}

\begin{document}

\begin{frontmatter} 

\title{Charge symmetry breaking in $\Lambda$ hypernuclei revisited} 

\author{Avraham Gal}
\address{Racah Institute of Physics, The Hebrew University, Jerusalem 
91904, Israel} 

\date{\today} 

\begin{abstract} 
The large charge symmetry breaking (CSB) implied by the $\Lambda$ 
binding energy difference $\Delta B^{4}_{\Lambda}(0^+_{\rm g.s.})\equiv 
B_{\Lambda}$(\lamb{4}{He})$-$$B_{\Lambda}$(\lamb{4}{H})~$=0.35\pm 0.06$~MeV of 
the $A=4$ mirror hypernuclei ground states, determined from emulsion studies, 
has defied theoretical attempts to reproduce it in terms of CSB in hyperon 
masses and in hyperon-nucleon interactions, including one pion exchange 
arising from $\Lambda-\Sigma^0$ mixing. Using a schematic strong-interaction 
$\Lambda N\leftrightarrow\Sigma N$ coupling model developed by Akaishi and 
collaborators for $s$-shell $\Lambda$ hypernuclei, we revisit the evaluation 
of CSB in the $A=4$ $\Lambda$ hypernuclei and extend it to $p$-shell mirror 
$\Lambda$ hypernuclei. The model yields values of $\Delta B^{4}_{\Lambda}
(0^+_{\rm g.s.})\sim 0.25$~MeV. Smaller size and mostly negative $p$-shell 
binding energy differences are calculated for the $A=7-10$ mirror hypernuclei, 
in rough agreement with the few available data. CSB is found to reduce by 
almost 30~keV the 110~keV \lam{10}{B} g.s. doublet splitting anticipated 
from the hyperon-nucleon strong-interaction spin dependence, 
thereby explaining the persistent experimental failure to observe 
the $2^-_{\rm exc}\to 1^-_{\rm g.s.}$ $\gamma$-ray transition.  
\end{abstract} 


\end{frontmatter}

\section{Introduction} 
\label{sec:intro} 

Charge symmetry breaking (CSB) in nuclear physics is primarily identified 
by considering the difference between $nn$ and $pp$ scattering lengths, or 
the binding-energy difference between the mirror nuclei $^3$H and $^3$He 
\cite{Miller06}. In these nuclei, about 70 keV out of the Coulomb-dominated 
764~keV binding-energy difference is commonly attributed to CSB which can 
be explained either by $\rho^0\omega$ mixing in one-boson exchange models 
of the $NN$ interaction, or by considering $N\Delta$ intermediate-state mass 
differences in models limited to pseudoscalar meson exchanges \cite{Mach01}. 

\begin{figure}[htb] 
\begin{center} 
\includegraphics[width=0.9\textwidth]{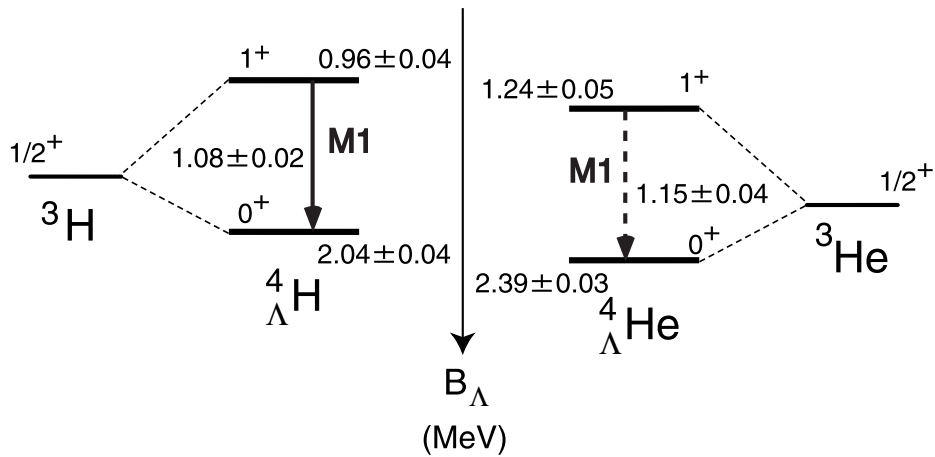} 
\caption{Level diagram of mirror hypernuclei (\lamb{4}{H},~\lamb{4}{He}) 
obtained by adding a $\Lambda$ hyperon to the mirror nuclei ($^3$H,~$^3$He). 
The $\Lambda$ separation energies, also loosely termed $\Lambda$ binding 
energies ($B_{\Lambda}$ in MeV), are taken from emulsion work \cite{Davis05}. 
Figure adapted from Ref.~\cite{Tamura13}.} 
\label{fig:A4} 
\end{center} 
\end{figure}

In $\Lambda$ hypernuclei, in contrast, CSB appears to be 
considerably stronger, judging by the binding-energy difference 
$\Delta B^{4}_{\Lambda}(0^+_{\rm g.s.})$=0.35$\pm$0.06~MeV 
deduced from the level diagram of the mirror hypernuclei 
(\lamb{4}{H},~\lamb{4}{He}) in Fig.~\ref{fig:A4}. A very recent measurement 
of \lamb{4}{H}$\to$$^4$He+$\pi^-$ decay at MAMI \cite{MAMI15} reduces $\Delta 
B^{4}_{\Lambda}(0^+_{\rm g.s.})$ to 0.27$\pm$0.10~MeV, consistent with its 
emulsion value \cite{Davis05}. Fig.~\ref{fig:A4} also suggests that $\Delta 
B^{4}_{\Lambda}(1^+_{\rm exc})$ is almost as large as $\Delta B^{4}_{\Lambda}
(0^+_{\rm g.s.})$. However, the deduction of the $1^+$ excitation energy 
in \lamb{4}{He} from the 1.15~MeV $1^+_{\rm exc}\to 0^+_{\rm g.s.}$ 
$\gamma$-ray transition \cite{Bedj79} is not as firm as the one for 
\lamb{4}{H} \cite{Tamura13}. In passing we mention the weak 1.42~MeV 
$\gamma$-ray transition reported in Ref.~\cite{Bamb73} that would imply 
almost no CSB splitting of the $1^+_{\rm exc}$ states if its assignment 
to \lamb{4}{He} gets confirmed.  

The large $\Delta B^{4}_{\Lambda}$ values reported for both $0^+_{\rm g.s.}$ 
and $1^+_{\rm exc}$ states have defied theoretical attempts to explain these 
differences in terms of hadronic or quark CSB mechanisms within four-body 
calculations \cite{Carlson91,Coon99,Nogga02,Nogga07,Nogga13}. Meson mixing, 
including $\rho^0\omega$ mixing which explains CSB in the $A=3$ nuclei, gives 
only small {\it negative} contributions about $-30$ and $-10$~keV for $\Delta 
B^{4}_{\Lambda}(0^+_{\rm g.s.})$ and $\Delta B^{4}_{\Lambda}(1^+_{\rm exc})$, 
respectively \cite{Coon99}. CSB contributions to $\Delta B^{4}_{\Lambda}
(0^+_{\rm g.s.})$ from one- and two-pion exchange interactions in $YNNN$ 
coupled-channel calculations \cite{Nogga02,Nogga07} with hyperons $Y=\Lambda,
\Sigma$ amount to as much as 100~keV; this holds for the OBE-based Nijmegen 
NSC97 models \cite{NSC97} which are widely used in $\Lambda$-hypernuclear 
structure calculations.{\footnote{Contradictory statements were made in 
Refs.~\cite{Carlson91,Nogga02} on the ability of the earlier Nijmegen model 
NSC89 \cite{NSC89} to reproduce $\Delta B^{4}_{\Lambda}(0^+_{\rm g.s.})$. 
We note that the strength of CSB contributions in this work, on p.~2236, 
is inflated erroneously by a factor $g_{NNM}$, about 3.7 for pion exchange, 
which might have propagated into some of the calculations claiming to have 
resolved the CSB puzzle for $\Delta B^{4}_{\Lambda}(0^+_{\rm g.s.})$ using 
NSC89.}} 

Binding energies of ground states in $p$-shell mirror $\Lambda$ hypernuclei, 
determined from emulsion studies \cite{Davis05}, suggest much weaker CSB 
effects than for $A=4$, with values of $\Delta B_{\Lambda}^{A} \equiv 
B_{\Lambda}(A,I,I_z)-B_{\Lambda}(A,I,-I_z)$ ($I_z>0$) consistent with 
zero for $A=8$ and somewhat negative beyond \cite{Davis05}. Accommodating 
$\Delta B_{\Lambda}$ values in the $p$ shell with $\Delta B^{4}_{\Lambda}$ 
by using reasonable phenomenological CSB interactions is impossible, 
as demonstrated in recent four-body cluster-model calculations of $p$-shell 
$\Lambda$ hypernuclei~\cite{Hiyama13}. This difficulty may be connected to 
the absence of explicit $\Lambda N\leftrightarrow \Sigma N$ coupling in these 
cluster-model calculations, given that such explicit coupling was shown to 
generate non-negligible CSB contributions to $\Delta B^{4}_{\Lambda}(0^+_{\rm 
g.s.})$ \cite{Nogga13}. It is our purpose in this note to use a schematic 
$\Lambda N\leftrightarrow\Sigma N$ coupling model, proposed by Akaishi et 
al.~\cite{Akaishi00,Akaishi02} for $s$-shell $\Lambda$ hypernuclei and 
extended by Millener~\cite{Millener08} to the $p$ shell, for calculating 
$\Delta B_{\Lambda}$ values in both $s$ and $p$ shells, thereby making 
predictions on CSB effects in $p$-shell $\Lambda$ hypernuclei consistently 
with a relatively sizable value of $\Delta B^{4}_{\Lambda}(0^+_{\rm g.s.})$. 

The paper is organized as follows. In Sect.~\ref{sec:CSB}, we update 
the original treatment by Dalitz and Von Hippel \cite{DvH64} of the 
$\Lambda-\Sigma^0$ mixing mechanism for generating CSB one-pion exchange 
contributions in $\Lambda$ hypernuclei, linking it to the strong-interaction 
$\Lambda N\leftrightarrow \Sigma N$ coupling model employed in this 
work. Our CSB calculations for the $A=4$ hypernuclei are sketched in 
Sect.~\ref{sec:s-shell} and their results are compared with those reported in 
several $YNNN$ four-body calculations \cite{Coon99,Nogga02,Nogga07,Nogga13}. 
Finally, CSB contributions in $p$-shell mirror $\Lambda$ hypernuclei, 
evaluated here for the first time, are reported in Sect.~\ref{sec:p-shell}.

\section{Pionic CSB contributions in $\Lambda$ hypernuclei} 
\label{sec:CSB} 

The $I=0$ isoscalar nature of the $\Lambda$ hyperon forbids it to 
emit or absorb a single pion, and hence there is no one-pion exchange 
(OPE) contribution to the $\Lambda N$ strong interaction. However, 
by allowing for $\Lambda-\Sigma^0$ mixing in SU(3), a CSB OPE contribution 
arises \cite{DvH64} with $\Lambda\Lambda\pi$ coupling constant 
\begin{equation} 
g_{\Lambda\Lambda\pi}=-2\;\frac{\langle\Sigma^0|\delta M|\Lambda\rangle}
{M_{\Sigma^0}-M_{\Lambda}}g_{\Lambda\Sigma\pi}=-0.0297g_{\Lambda\Sigma\pi}, 
\label{eq:DvH} 
\end{equation} 
where the matrix element of the mass mixing operator $\delta M$ is given by 
\begin{equation} 
\langle\Sigma^0|\delta M|\Lambda\rangle=\frac{1}{\sqrt 3}
(M_{\Sigma^0}-M_{\Sigma^+}+M_p-M_n).   
\label{eq:deltaM} 
\end{equation} 
The resulting CSB OPE potential is given by 
\begin{equation} 
V_{\Lambda N}^{\rm CSB}({\rm OPE})=-0.0242\,\tau_{Nz}\frac{f_{NN\pi}^2}{4\pi}
\frac{m_{\pi}}{3}[{\vec\sigma}_{\Lambda}\cdot{\vec\sigma}_N + T(r)
S(\hat{r};\,{\vec\sigma}_{\Lambda},{\vec\sigma}_N)]Y(r), 
\label{eq:OPE} 
\end{equation} 
where the z component of the isospin Pauli matrix ${\vec\tau}_N$ assumes 
the values $\tau_{Nz}=\pm 1$ on protons and neutrons, respectively, 
$Y(r)=\exp(-m_{\pi}r)/(m_{\pi}r)$ is a Yukawa form, and the tensor 
contribution is specified by 
\begin{equation} 
T(r)=1+\frac{3}{m_{\pi}r}+\frac{3}{(m_{\pi}r)^2}, \;\;\; S(\hat{r};\,
{\vec\sigma}_{\Lambda},{\vec\sigma}_N)=3{\vec\sigma}_{\Lambda}\cdot \hat{r}\; 
{\vec\sigma}_N\cdot \hat{r}-{\vec\sigma}_{\Lambda}\cdot{\vec\sigma}_N. 
\label{eq:Yukawa} 
\end{equation} 
In Eq.~(\ref{eq:OPE}), the transition $g_{\Lambda\Sigma\pi}\to f_{NN\pi}$ 
was made in accordance with NSC models, using $f_{NN\pi}^2/4\pi=0.0740$ 
from NSC89, Table IV in Ref~\cite{NSC89}. 

Since Pauli-spin $S_{pp}=0$ and $S_{nn}=0$ hold in 
\lamb{4}{He}($0^+_{\rm g.s.}$) and in \lamb{4}{H}($0^+_{\rm g.s.}$), 
respectively, the CSB potential (\ref{eq:OPE}) which is linear in the 
nucleon spin gives no contribution from these same-charge nucleons. 
Therefore $\tau_{Nz}=\mp 1$ owing to the odd nucleon in these hypernuclei, 
respectively, and since ${\vec\sigma}_{\Lambda}\cdot{\vec\sigma}_N=-3$ 
holds for this odd nucleon, one gets a {\it positive} $\Delta B_{\Lambda}$ 
contribution from the central spin-spin part which provides the only 
nonvanishing contribution for simple $L=0$ wavefunctions. For the 
$0^+_{\rm g.s.}$ wavefunction used by Dalitz and Von Hippel \cite{DvH64}, 
and updating the values of the coupling constants used in their work to 
those used here, one gets $\Delta B_{\Lambda}^{\rm OPE}(0^+_{\rm g.s.})
\approx 95$~keV, a substantial single contribution with respect to 
$\Delta B_{\Lambda}^{\rm exp}(0^+_{\rm g.s.})=350\pm 60$~keV. 

The $\Lambda-\Sigma^0$ mixing mechanism gives rise also to a variety of (e.g. 
$\rho$) meson exchanges other than OPE. In baryon-baryon models that consider 
{\it explicitly} the strong-interaction $\Lambda N\leftrightarrow\Sigma N$ 
coupling, the matrix element of $V_{\Lambda N}^{\rm CSB}$ is related to 
a suitably chosen strong-interaction isospin $I_{NY}=1/2$ matrix element 
$\langle N\Sigma|V|N\Lambda\rangle$ by generalizing Eq.~(\ref{eq:DvH}): 
\begin{equation} 
\langle N\Lambda|V_{\Lambda N}^{\rm CSB}|N\Lambda\rangle = 
-0.0297\,\tau_{Nz}\,\frac{1}{\sqrt{3}}\,\langle N\Sigma|V|N\Lambda\rangle , 
\label{eq:OME} 
\end{equation}  
where the isospin Clebsch-Gordan coefficient $1/\sqrt{3}$ accounts for the 
$N\Sigma^0$ amplitude in the $I_{NY}=1/2$ $N\Sigma$ state, and the space-spin 
structure of this $N\Sigma$ state is taken identical with that of the 
$N\Lambda$ state sandwiching $V_{\Lambda N}^{\rm CSB}$.

\section{CSB in the $A=4$ hypernuclei} 
\label{sec:s-shell} 

\begin{table}[htb]
\begin{center}
\caption{$\Lambda\Sigma$ $s$-shell matrix elements 
${\bar V}^{0s}_{\Lambda\Sigma}$ and $\Delta^{0s}_{\Lambda\Sigma}$, 
downward energy shifts $\delta E_{\downarrow}(J^{\pi})$ calculated 
for the $A=4$ hypernuclear states, the resulting $\Delta 
E_{\Lambda\Sigma}(0^+_{\rm g.s.}-1^+_{\rm exc})$ in models 
$(\Lambda\Sigma)_{\rm e,f}$ \cite{Millener08} and the full excitation energy 
$\Delta E(0^+_{\rm g.s.}-1^+_{\rm exc})$ \cite{Nogga02,Akaishi00,Akaishi02}, 
all in MeV.} 
\begin{tabular}{ccccccccc} 
\hline 
NSC97 & ${\bar V}^{0s}_{\Lambda\Sigma}$ & $\Delta^{0s}_{\Lambda\Sigma}$ & 
$\delta E_{\downarrow}(0^+_{\rm g.s.})$&$\delta E_{\downarrow}(1^+_{\rm exc})$ 
& $\Delta E_{\Lambda\Sigma}$ & \multicolumn{3}{c}
{$\Delta E(0^+_{\rm g.s.}-1^+_{\rm exc})$} \\  model & \multicolumn{5}{c}{as 
calculated in models $(\Lambda\Sigma)_{\rm e,f}$ \cite{Millener08}} & 
\cite{Nogga02} & \cite{Akaishi00} & \cite{Akaishi02} \\ 
\hline 
NSC97$_{\rm e}$ & 2.96 & 5.09 & 0.574 & 0.036 & 0.539 & 0.75 & 0.89 
& 1.13 \\ 
NSC97$_{\rm f}$ & 3.35 & 5.76 & 0.735 & 0.046 & 0.689 & 1.10 & 1.48 
& 1.51 \\ 
\hline 
\end{tabular} 
\label{tab:A=4} 
\end{center} 
\end{table} 
Following hyperon-core calculations of $s$-shell $\Lambda$ hypernuclei by 
Akaishi et al.~\cite{Akaishi00} we use $G$-matrix $YN$ effective interactions 
derived from NSC97 models to calculate CSB contributions from 
Eq.~(\ref{eq:OME}). 
%These $0s_N0s_{Y}$ effective interactions were 
%subsequently extended by providing $0p_N0s_{Y}$ shell-model matrix elements 
%which are consistent with the role $\Lambda N\leftrightarrow\Sigma N$ 
%($\Lambda\Sigma$) coupling appears to play in a shell-model reproduction 
%of hypernuclear $\gamma$-ray transition energies \cite{Millener12}. For 
%a recent application to neutron-rich hypernuclei, see Ref.~\cite{Galmil13}. 
The $\Lambda\Sigma$ $0s_N0s_{Y}$ effective interaction $V_{\Lambda\Sigma}$ 
is given in terms of a spin-dependent central interaction  
\begin{equation} 
V_{\Lambda\Sigma}=({\bar V}_{\Lambda\Sigma}+\Delta_{\Lambda\Sigma}{\vec s}_N
\cdot{\vec s}_{Y})\sqrt{4/3}\;{\vec t}_N\cdot{\vec t}_{\Lambda\Sigma}, 
\label{eq:V_YN} 
\end{equation}     
where ${\vec t}_{\Lambda\Sigma}$ converts a $\Lambda$ to $\Sigma$ in isospace. 
The $s$-shell matrix elements ${\bar V}^{0s}_{\Lambda\Sigma}$ and $\Delta^{
0s}_{\Lambda\Sigma}$ are listed in Table~\ref{tab:A=4} for two such $G$-matrix 
models denoted $(\Lambda\Sigma)_{\rm e,f}$. Also listed are the calculated 
downward energy shifts $\delta E_{\downarrow}(J^{\pi})$ defined by $\delta E_{
\downarrow}(J^{\pi})= v^2(J^{\pi})/(80~{\rm MeV})$, where the $0s_N^30s_Y$ 
matrix elements $v(J^{\pi})$ for $A=4$ are given in terms of $\Lambda\Sigma$ 
two-body matrix elements by 
\begin{equation} 
v(0^+_{\rm g.s.})={\bar V}^{0s}_{\Lambda\Sigma}+\frac{3}{4}
\Delta^{0s}_{\Lambda\Sigma},\;\;\; v(1^+_{\rm exc})={\bar V}^{0s}_{\Lambda
\Sigma}-\frac{1}{4}\Delta^{0s}_{\Lambda\Sigma}. 
\label{eq:v(JP)} 
\end{equation} 
We note that the diagonal $0s_N0s_{\Sigma}$ interaction matrix elements have 
little effect in this coupled-channel model because of the large energy 
denominators of order $M_{\Sigma}-M_{\Lambda}\approx 80$~MeV with which they 
appear. Finally, by listing $\Delta E_{\Lambda\Sigma}(0^+_{\rm g.s.} - 1^+_{
\rm exc})$ from Refs.~\cite{Akaishi00,Akaishi02} we demonstrate the sizable 
contribution of $\Lambda\Sigma$ coupling to the excitation energy $\Delta 
E(0^+_{\rm g.s.} - 1^+_{\rm exc})\approx 1.1$~MeV deduced from the 
$\gamma$-ray transition energies marked in Fig.~\ref{fig:A4}. 
For comparison, the full $\Delta E(0^+_{\rm g.s.}-1^+_{\rm exc})$ in these 
$(\Lambda\Sigma)_{\rm e,f}$ models, and as calculated by Nogga~\cite{Nogga02} 
using the underlying Nijmegen models NSC97$_{\rm e,f}$, are also listed in 
the table. 

Having discussed the effect of strong-interaction $\Lambda\Sigma$ coupling, 
we now discuss the CSB splittings $\Delta B^{4}_{\Lambda}(0^+_{\rm g.s.})$ 
and $\Delta B^{4}_{\Lambda}(1^+_{\rm exc})$. Results of our $\Lambda\Sigma$ 
coupling model calculations, using Eq.~(\ref{eq:OME}) for one of several 
contributions, are listed in the last two lines of Table~\ref{tab:CSB}, 
preceded by results obtained in other models within genuine four-body 
calculations \cite{Coon99,Nogga02,Nogga07,Nogga13}. Partial contributions to 
$\Delta B^{4}_{\Lambda}(0^+_{\rm g.s.})$ are listed in columns 2--5, whereas 
for $\Delta B^{4}_{\Lambda}(1^+_{\rm exc})$ only its total value is listed.  

\begin{table}[htb]
\begin{center}
\caption{Calculated CSB contributions to $\Delta B^{4}_{\Lambda}(0^+_{
\rm g.s.})$ and total values of $\Delta B^{4}_{\Lambda}(0^+_{\rm g.s.})$ 
and $\Delta B^{4}_{\Lambda}(1^+_{\rm exc})$, in keV, from several model 
calculations of the $A=4$ hypernuclei. Recall that $\Delta B_{\Lambda}^{
\rm exp}(0^+_{\rm g.s.})=350\pm 60$~keV \cite{Davis05}.} 
\begin{tabular}{lccccrcr} 
\hline 
\lamb{4}{He}--\lamb{4}{H} & $P_{\Sigma}(\%)$ & $\Delta T_{YN}$ & $\Delta V_C$ 
& $\Delta V_{YN}$ & $\Delta B^{4}_{\Lambda}$ & & $\Delta B^{4}_{\Lambda}$ \\  
model & $0^+_{\rm g.s.}$ & $0^+_{\rm g.s.}$ & $0^+_{\rm g.s.}$ & 
$0^+_{\rm g.s.}$ & $0^+_{\rm g.s.}$ & & $1^+_{\rm exc}$ \\  
\hline 
$\Lambda NNN$ \cite{Coon99} & -- & -- & $-$42 & 91 & 49 & & $-$61 \\  
NSC97$_{\rm e}$ \cite{Nogga02} & 1.6 & 47 & $-$16 & 44 & 75 & & $-$10 \\ 
NSC97$_{\rm f}$ \cite{Nogga07} & 1.8 &    &       &    & 100 & & $-$10 \\  
NLO chiral \cite{Nogga13} & 2.1 & 55 & $-$9 & -- & 46 & &  \\ 
$(\Lambda\Sigma)_{\rm e}$ [present] & 0.72 & 39 & $-$45 & 232 & 226 & & 30 \\
$(\Lambda\Sigma)_{\rm f}$ [present] & 0.92 & 49 & $-$46 & 263 & 266 & & 39 \\ 
\hline 
\end{tabular} 
\label{tab:CSB} 
\end{center} 
\end{table} 

All of the models listed in Table~\ref{tab:CSB} except for \cite{Coon99} 
include $\Lambda\Sigma$ coupling, with $0^+_{\rm g.s.}$ $\Sigma NNN$ 
admixture probabilities $P_{\Sigma}\approx P_{\Sigma^{\pm}}+P_{\Sigma^0}$ 
in (\lamb{4}{He},~\lamb{4}{H}) respectively, and $P_{\Sigma^{\pm}}\approx 
\frac{2}{3}P_{\Sigma}$. The $1^+_{\rm exc}$ $\Sigma NNN$ admixtures (unlisted) 
are considerably weaker than the listed $0^+_{\rm g.s.}$ admixtures.
Charge asymmetric kinetic-energy contributions to $\Delta B_{\Lambda}$, 
dominated by $\Sigma N$ intermediate-state mass differences, are marked 
$\Delta T_{YN}$ in the table. In the present $\Lambda\Sigma$ coupling model 
these are given for the $0^+_{\rm g.s.}$ by \cite{Nogga02} 
\begin{equation} 
\Delta T_{YN}(0^+_{\rm g.s.})\approx\frac{2}{3}\,P_{\Sigma}\,
(M_{\Sigma^-}-M_{\Sigma^+}), 
\label{eq:kin} 
\end{equation} 
yielding as much as 50 keV, in agreement with those four-body calculations 
where such mass differences were introduced \cite{Nogga02,Nogga07,Nogga13}. 
The next column in the table, $\Delta V_C=\Delta V_C^{\Lambda}+\Delta 
V_C^{\Sigma}$, addresses contributions arising from nuclear-core Coulomb 
energy modifications induced by the hyperons. $\Delta V_C^{\Lambda}$ is 
negative, its size ranges from less than 10~keV \cite{Nogga02,Nogga13} 
to about 40~keV \cite{Coon99}. $\Delta V_C^{\Sigma}$ which accounts for 
$\Sigma^{\pm}p$ Coulomb energies in the $\Sigma NNN$ admixed components 
is also negative and uniformly small with size of a few keV at most. 
The values assigned to $\Delta V_C$ in the $\Lambda\Sigma$ model use 
values from Ref.~\cite{Coon99} for $\Delta V_C^{\Lambda}$ and the estimate 
$\Delta V_C^{\Sigma}\approx -\frac{2}{3}P_{\Sigma}E_C(^3{\rm He})$ for 
$\Delta V_C^{\Sigma}$, where $E_C(^3{\rm He})=644$~keV is the Coulomb energy 
of $^3$He.  

The next contribution, $\Delta V_{YN}$, is derived from 
$V_{\Lambda N}^{\rm CSB}$. No $\Delta V_{YN}$ contributions are available 
from the coupled channels calculation by Hiyama et al.~\cite{Hiyama02} 
(not listed here) and also from the recent chiral-model calculation in which 
CSB contributions are disregarded \cite{Nogga13} in order to remain consistent 
with EFT power counting rules that exclude CSB from the NLO chiral version 
of the $YN$ interaction \cite{NLO13}. With the exception of the purely 
$\Lambda NNN$ four-body calculation of Ref.~\cite{Coon99}, all those 
models for which a nonzero value is listed in the table effectively used 
Eq.~(\ref{eq:OME}) to evaluate $\Delta V_{YN}(0^+_{\rm g.s.})$. This ensures 
that meson exchanges arising from $\Lambda-\Sigma^0$ mixing beyond OPE 
are also included in the calculated CSB contribution. Generally, the CSB 
potential contribution $\Delta V_{YN}(0^+_{\rm g.s.})$ is not linked in 
any simple model-independent way to the $\Sigma$ admixture probability 
$P_{\Sigma}(0^+_{\rm g.s.})$. For example, the calculations using 
NSC97 \cite{Nogga02,Nogga07} produce too little CSB contributions, 
whereas the present $\Lambda\Sigma$ model, in spite of its weaker $\Sigma$ 
admixtures, gives sizable contributions which essentially resolve the CSB 
puzzle in the $0^+_{\rm g.s.}$ of the $A=4$ hypernuclei. Indeed, using 
a typical $\Lambda\Sigma$ strong-interaction matrix element $\langle N
\Sigma|V(0^+_{\rm g.s.})|N\Lambda\rangle\sim 7$~MeV in Eq.~(\ref{eq:OME}) 
one obtains $P_{\Sigma}(0^+_{\rm g.s.})=0.77\%$ and a CSB contribution of 
240~keV to $\Delta B_{\Lambda}(0^+_{\rm g.s.})$; this CSB contribution is 
proportional to $\sqrt{P_{\Sigma}}$ in the present $\Lambda\Sigma$ model. 

The resulting values of $\Delta B^{4}_{\Lambda}(0^+_{\rm g.s.})$ listed in 
Table~\ref{tab:CSB} are smaller than 100~keV within the calculations presented 
in Refs.~\cite{Coon99,Nogga02,Nogga07,Nogga13}, leaving the $A=4$ CSB puzzle 
unresolved, while being larger than 200~keV in the present $\Lambda\Sigma$ 
model and thereby getting considerably closer to the experimentally reported 
$0^+_{\rm g.s.}$ CSB splitting. The main difference between these two groups 
of calculations arises from the difference in the CSB potential contributions 
$\Delta V_{YN}(0^+_{\rm g.s.})$. A similarly large difference also appears 
between the CSB potential {\it negative} contributions $\Delta V_{YN}
(1^+_{\rm exc})$ in the calculations of Refs.~\cite{Coon99,Nogga02,Nogga07} 
and the {\it positive} contributions $\Delta V_{YN}(1^+_{\rm exc})$ in
the present $\Lambda\Sigma$ model, resulting in large but different 
$\Delta B^{4}_{\Lambda}(0^+_{\rm g.s.})-\Delta B^{4}_{\Lambda}(1^+_{\rm exc})$ 
values, about 200~keV in the present $\Lambda\Sigma$ model and about 100~keV 
for all other calculations \cite{Coon99,Nogga02,Nogga07}. A common feature of  
all CSB model calculations so far is that none of them is able to generate 
values in excess of 50~keV for $\Delta B^{4}_{\Lambda}(1^+_{\rm exc})$. 

A direct comparison between the NCS97 models and the present $\Lambda\Sigma$ 
model is not straightforward because the $\Lambda\Sigma$ coupling 
in NSC97 models is dominated by tensor components, whereas no tensor 
components appear in present $\Lambda\Sigma$ model. It is worth noting, 
however, that the $\rho$ exchange contribution to the matrix element 
$\langle N\Sigma|V|N\Lambda\rangle$ in Eq.~(\ref{eq:OME}) is of opposite 
sign to that of OPE for the tensor $\Lambda\Sigma$ coupling which dominates 
in NSC models, leading to cancellations, whereas both $\rho$ exchange and OPE 
contribute constructively in the present central $\Lambda\Sigma$ coupling 
model in agreement with the calculation by Coon et al.~\cite{Coon99} which 
also has no tensor components.{\footnote{It is worth noting that the $\rho$ 
exchange CSB contribution calculated in Ref.~\cite{Coon99} is of the same sign 
and remarkably stronger than the OPE CSB contribution.}} This point deserves 
further study by modeling various input $YN$ interactions in future four-body 
calculations.

\section{CSB in $p$-shell hypernuclei} 
\label{sec:p-shell} 

Several few-body cluster-model calculations, of the $A=7$, $I=1$ isotriplet 
\cite{Hiyama09} and the $A=10$, $I=\frac{1}{2}$ isodoublet \cite{Hiyama12}, 
have considered the issue of CSB contributions to $\Lambda$ binding energy 
differences of $p$-shell mirror hypernuclei. It was verified in these 
calculations that the introduction of a $\Lambda N$ phenomenological CSB 
interaction fitted to $\Delta B^{4}_{\Lambda}$, for both $0^+_{\rm g.s.}$ 
and $1^+_{\rm exc}$ states, failed to reproduce the observed $\Delta B^{A}_{
\Lambda}$ values in these $p$-shell hypernuclei; in fact, it only aggravated 
the discrepancy between experiment and calculations. Although it is possible 
to reproduce the observed values by introducing additional CSB components 
that hardly affect $\Delta B^{4}_{\Lambda}$, this prescription lacks any 
physical origin and is therefore questionable, as acknowledged very recently 
by Hiyama \cite{Hiyama13}. Here we explore $p$-shell CSB contributions, 
extending the NSC97e model $0s_N0s_{Y}$ effective interactions considered 
in Sect.~\ref{sec:s-shell}, by providing $(\Lambda\Sigma)_{\rm e}$ $0p_N0s_Y$ 
central-interaction matrix elements which are consistent with the role 
$\Lambda N\leftrightarrow\Sigma N$ coupling appears to play in a shell-model 
reproduction of hypernuclear $\gamma$-ray transition energies 
\cite{Millener12}: 
\begin{equation} 
{\bar V}^{0p}_{\Lambda\Sigma}=1.45, \;\;\;\;\;\;\;\;\;\;
\Delta^{0p}_{\Lambda\Sigma}=3.04 \;\;\;\;\;\;\;\;\;\; {\rm (in~MeV)}. 
\label{eq:0p} 
\end{equation} 
These $p$-shell matrix elements are smaller by roughly a factor of two from 
the corresponding $s$-shell matrix elements in Table~\ref{tab:A=4}, reflecting 
a reduced weight, about 1/2, with which the dominant relative $s$-wave matrix 
elements of $V_{NY}$ appear in the $p$ shell. This suggests that $\Sigma$ 
admixtures which are quadratic in these matrix elements, are weaker roughly 
by a factor of 4 with respect to the $s$-shell calculation, and also smaller 
CSB interaction contributions in the $p$ shell with respect to those in the 
$A=4$ hypernuclei, although only by a factor of 2. To evaluate these CSB 
contributions, instead of applying the one-nucleon or nucleon-hole expression 
(\ref{eq:OME}) valid in the $s$ shell, we use in the $p$ shell the general 
multi-nucleon expression for $V_{\Lambda N}^{\rm CSB}$ obtained by summing 
over $p$-shell nucleons:  
\begin{equation} 
V_{\Lambda N}^{\rm CSB} = -0.0297\,\frac{1}{\sqrt{3}}
\sum_j{({\bar V}^{0p}_{\Lambda\Sigma}+\Delta^{0p}_{\Lambda\Sigma}
{\vec s}_j\cdot{\vec s}_Y)\tau_{jz}}. 
\label{eq:VCSB} 
\end{equation} 

Results of applying the present $(\Lambda\Sigma)_{\rm e}$ coupling model 
to several pairs of g.s. levels in $p$-shell hypernuclear isomultiplets 
are given in Table~\ref{tab:pshell}. All pairs except for $A=7$ are mirror 
hypernuclei identified in emulsion \cite{Davis05} where binding energy 
systematic uncertainties are largely canceled out in forming the listed 
$\Delta B_{\Lambda}^{\rm exp}$ values. For $A=7$ we compensated for the 
unavailability of a reliable $B_{\Lambda}$(\lamb{7}{He}) value from 
emulsion by replacing it with $B_{\Lambda}$(\lamb{7}{Li}$^{\ast}$), 
established by observing the 3.88~MeV $\gamma$-ray transition 
\lamb{7}{Li}$^{\ast}\to\gamma$+\lamb{7}{Li} \cite{Tamura00}. Recent 
$B_{\Lambda}$ values determined in electroproduction experiments at JLab 
for \lamb{7}{He} \cite{JLab13,Gogami14}, \lamb{9}{Li} \cite{JLab14} and 
\lam{10}{Be} \cite{Gogami14} were not used for lack of similar data on 
their mirror partners. 

\begin{table}[htb]
\begin{center}
\caption{CSB contributions to $\Delta B^{\rm calc}_{\Lambda}(\rm g.s.)$ values 
in $p$-shell hypernuclear isomultiplets, using the $(\Lambda\Sigma)_{\rm e}$ 
coupling model. The $s$-shell contributions to $\Delta B^{4}_{\Lambda}
(0^+_{\rm g.s.})$ from Table~\ref{tab:CSB} are also listed for comparison.} 
\begin{tabular}{cclccrrr} 
\hline 
\lamb{A}{Z$_{>}$}--\lamb{A}{Z$_{<}$} & $I,J^{\pi}$ & 
$P_{\Sigma}$ & $\Delta T_{YN}$ & $\Delta V_C$ & $\Delta V_{YN}$ & $\Delta 
B_{\Lambda}^{\rm calc}$ & $\Delta B_{\Lambda}^{\rm exp}$\cite{Davis05} \\ 
pairs & & (\%) & (keV) & (keV) & (keV) & (keV) & (keV) \\ 
\hline 
\lamb{4}{He}--\lamb{4}{H} & $\frac{1}{2},0^+$ & 0.72 & 39 & $-$45 & 232 & 
226 & $+$350$\pm$60 \\
\hline 
\lamb{7}{Be}--\lamb{7}{Li}$^{\ast}$ & $1,{\frac{1}{2}}^+$ & 0.12 & 3 & 
$-$70 \cite{Hiyama09} & 50 & $-$17 & $-$100$\pm$90 \\ 
\lamb{8}{Be}--\lamb{8}{Li} & $\frac{1}{2},1^-$ & 0.20 & 11 & 
$-$81 \cite{Hiyama02a} & 119 & $+$49 & $+$40$\pm$60 \\ 
\lamb{9}{B}--\lamb{9}{Li} & $1,{\frac{3}{2}}^+$ & 0.23 & 10 & 
$-$145 \cite{Millener15} & 81 & $-$54 & $-$210$\pm$220 \\ 
\lam{10}{B}--\lam{10}{Be} & $\frac{1}{2},1^-$ & 0.053 & 3 & 
$-$156 \cite{Millener15} & 17 & $-$136 & $-$220$\pm$250 \\ 
\hline 
\end{tabular} 
\label{tab:pshell} 
\end{center} 
\end{table} 

The $\Sigma$ admixture percentages $P_{\Sigma}$ in Table~\ref{tab:pshell} 
follow from $\Lambda\Sigma$ strong-interaction contributions to $p$-shell 
hypernuclear g.s. energies computed in Ref.~\cite{Millener12}, and the 
associated CSB kinetic-energy contributions $\Delta T_{YN}$ were calculated 
using a straightforward generalization of Eq.~(\ref{eq:kin}). These 
contributions, of order 10~keV and less, are considerably weaker than 
the $\Delta T_{YN}$ contributions to $\Delta B^{4}_{\Lambda}$ listed in 
Table~\ref{tab:CSB}, reflecting weaker $\Sigma$ admixtures in the $p$ shell 
as discussed following Eq.~(\ref{eq:0p}). 
The Coulomb-induced contributions $\Delta V_C$ are dominated by their $\Delta 
V^{\Lambda}_C$ components which were taken from Hiyama's cluster-model 
calculations \cite{Hiyama09,Hiyama02a} for $A=7,8$ and from Millener's 
shell-model calculations \cite{Millener15} for $A=9,10$. The shell-model 
estimate of $-$156~keV adopted here for $A=10$ is somewhat smaller than the 
$-$180~keV cluster-model result \cite{Hiyama12}. The $\Delta V^{\Sigma}_C$ 
components are negligible, with size of 1~keV at most (for $A=8,9$). 
$\Delta V^{\Lambda}_C$ is always negative, as expected from the increased 
Coulomb repulsion owing to the increased proton separation energy in the 
$\Lambda$ hypernucleus with respect to its core. The sizable negative 
$p$-shell $\Delta V_C$ contributions, in distinction from their secondary 
role in forming the total $\Delta B^{4}_{\Lambda}(0^+_{\rm g.s.})$, exceed 
in size the positive $p$-shell $\Delta V_{YN}$ contributions by a large 
margin beginning with $A=9$, thereby resulting in clearly negative values 
of $\Delta B^{A}_{\Lambda}(\rm g.s.)$. 

The CSB $\Delta V_{YN}$ contributions listed in Table~\ref{tab:pshell} 
were calculated using weak-coupling $\Lambda$-hypernuclear shell-model 
wavefunctions in terms of the corresponding nuclear-core g.s. leading 
SU(4) supermultiplet components, except for $A=8$ where the first excited 
nuclear-core level had to be included. This proved to be a sound and useful 
approximation, yielding $\Lambda\Sigma$ strong-interaction contributions 
close to those given in Figs.~1--3 of Ref.~\cite{Millener12}.{\footnote{I 
am indebted to John Millener for providing me with some of the wavefunctions 
required here.}} Details will be given elsewhere. The listed $A=7-10$ values 
of $\Delta V_{YN}$ exhibit strong SU(4) correlations, marked in particular 
by the enhanced value of 119~keV for the SU(4) nucleon-hole configuration in 
\lamb{8}{Be}--\lamb{8}{Li} with respect to the modest value of 17~keV for 
the SU(4) nucleon-particle configuration in \lam{10}{B}--\lam{10}{Be}. This 
enhancement follows from the relative magnitudes of the Fermi-like interaction 
term ${\bar V}^{0p}_{\Lambda\Sigma}$ and its Gamow-Teller partner term 
$\Delta^{0p}_{\Lambda\Sigma}$ in Eq.~(\ref{eq:0p}). Noting that both $A=4$ 
and $A=8$ mirror hypernuclei correspond to SU(4) nucleon-hole configuration,
the roughly factor two ratio of $\Delta V_{YN}(A=4)=232$~keV to $\Delta 
V_{YN}(A=8)=119$~keV reflects the approximate factor of two for the ratio  
between $s$-shell to $p$-shell $\Lambda\Sigma$ matrix elements, as discussed 
following Eq.~(\ref{eq:0p}). 

Comparing $\Delta B_{\Lambda}^{\rm calc}$ with $\Delta B_{\Lambda}^{\rm exp}$ 
in Table~\ref{tab:pshell}, we note the reasonable agreement reached between 
the present $(\Lambda\Sigma)_{\rm e}$ coupling model calculation and 
experiment for all four pairs of $p$-shell hypernuclei, $A=7-10$, considered 
in this work. Extrapolating to heavier hypernuclei, one might naively expect 
negative values of $\Delta B_{\Lambda}^{\rm calc}$, as suggested by the listed 
$A=9,10$ values. However, this rests on the assumption that the negative 
$\Delta V^{\Lambda}_C$ contribution remains as large upon increasing $A$ 
as it is in the beginning of the $p$ shell, which need not be the case. 
As nuclear cores beyond $A=9$ become more tightly bound, the $\Lambda$ 
hyperon is unlikely to compress these nuclear cores as much as it does in 
lighter hypernuclei, so that the additional Coulomb repulsion in \lam{12}{C}, 
for example, over that in \lam{12}{B}, while still negative, may not be 
sufficiently large to offset the attractive CSB contribution. In making 
this argument we rely on the expectation, based on SU(4) supermultiplet 
fragmentation patterns in the $p$ shell, that $\Delta V_{YN}$ does not 
exceed $\sim$100~keV. 

Before closing the discussion of CSB in $p$-shell hypernuclei, we wish to 
draw attention to the state dependence of CSB splittings, recalling the vast 
difference between the calculated $\Delta B_{\Lambda}^{4}(0^+_{\rm g.s.})$ 
and $\Delta B_{\Lambda}^{4}(1^+_{\rm exc})$ in the $s$ shell. 
In Table~\ref{tab:doublets} we list CSB contributions 
$\Delta E^{\rm CSB}_{\Lambda\Sigma}$ to several g.s. doublet excitation 
energies, as well the excitation energies $\Delta E^{\rm CS}$ calculated 
by Millener~\cite{Millener12} using charge symmetric (CS) YN spin-dependent 
interactions, including CS $\Lambda\Sigma$ contributions 
$\Delta E^{\rm CS}_{\Lambda\Sigma}$ (also listed). It is tacitly assumed that 
$\Delta V^{\Lambda}_C$ is state independent for the hypernuclear g.s. doublet 
members. As for the other, considerably smaller contributions, we checked that 
$\Delta V^{\Sigma}_C$ remains at the 1~keV level and that the difference 
between the appropriate $\Sigma$-dominated $\Delta T_{YN}$ values is less than 
10~keV. Under these circumstances, it is sufficient to limit the discussion 
to the state dependence of $\Delta V_{YN}$ alone, although the splittings 
$\Delta E^{\rm CSB}_{\Lambda\Sigma}$ listed in the table include these other 
tiny contributions. 

\begin{table}[htb]
\begin{center}
\caption{Ground-state doublet splittings in several $p$-shell hypernuclei: 
$\Delta E^{\rm exp}$, $\Delta E^{\rm CS}$ from \cite{Millener12} using CS 
spin-dependent $YN$ interactions that include CS $\Lambda\Sigma$ contributions 
$\Delta E^{\rm CS}_{\Lambda\Sigma}$, and CSB $\Lambda\Sigma$ contributions 
$\Delta E^{\rm CSB}_{\Lambda\Sigma}$ considered in the present 
work, all in keV.} 
\begin{tabular}{ccccccc} 
\hline 
\lamb{A}{Z} & $J^{\pi}_{\rm exc}$ & $J^{\pi}_{\rm g.s.}$ & 
$\Delta E^{\rm exp}$ & $\Delta E^{\rm CS}$~\cite{Millener12} & 
$\Delta E^{\rm CS}_{\Lambda\Sigma}$~\cite{Millener12} & 
$\Delta E^{\rm CSB}_{\Lambda\Sigma}$ \\  \hline 
\lamb{8}{Li} & $2^-$ & $1^-$ & 442$\pm$2 \cite{Chrien90} & 445 & 149 & 
$-$53.2 \\ 
\lamb{9}{Li} & ${\frac{5}{2}}^+$ & ${\frac{3}{2}}^+$ & 570$\pm$120 
\cite{JLab14} & 590 & 116 & $+$12.5 \\ 
\lam{10}{B} & $2^-$ & $1^-$ & $< 100$ \cite{Chrien90,Tamura05} & 110 & 
$-$10 & $-$26.5 \\ 
\hline 
\end{tabular} 
\label{tab:doublets} 
\end{center} 
\end{table} 

Inspection of Table~\ref{tab:doublets} reveals that whereas CSB contributions 
$\Delta E^{\rm CSB}_{\Lambda\Sigma}$ are negligible in \lamb{9}{Li}, with 
respect to both $\Delta E^{\rm CS}_{\Lambda\Sigma}$ and to the total CS 
splitting $\Delta E^{\rm CS}$, they need to be incorporated in re-evaluating 
the g.s. doublet splittings in \lamb{8}{Li} and in \lam{10}{B}. 
\begin{itemize} 
\item In \lamb{8}{Li}, these $\Delta E^{\rm CSB}_{\Lambda\Sigma}$ 
contributions spoil the perhaps fortuitous agreement between $\Delta E^{
\rm exp}$, deduced from a tentative assignment of a $\gamma$-ray transition 
observed in the $^{10}{\rm B}(K^-,\pi^-)$\lam{10}{B} reaction continuum 
spectrum \cite{Chrien90}, and $\Delta E^{\rm CS}$ evaluated using the 
$YN$ spin-dependent interaction parameters deduced from well identified 
$\gamma$-ray transitions in other hypernuclei. The 50~keV discrepancy arising 
from adding $\Delta E^{\rm CSB}_{\Lambda\Sigma}$ surpasses significantly 
the typical 20~keV theoretical uncertainty in fitting doublet splittings 
in $p$-shell hypernuclei (see Table 1, Ref.~\cite{Millener12}). 
\item 
The inclusion of $\Delta E^{\rm CSB}_{\Lambda\Sigma}$ in the calculated 
\lam{10}{B} g.s. doublet splitting helps solving the longstanding puzzle of 
not observing the $2^-_{\rm exc}\to 1^-_{\rm g.s.}$ $\gamma$-ray transition, 
thereby placing an upper limit of 100~keV on this transition energy 
\cite{Chrien90,Tamura05}. Including our CSB calculated contribution would 
indeed lower the expected transition energy from 110~keV to about 85~keV, 
in accordance with the experimental upper limit.{\footnote{Adding CSB is 
not essential in the \lam{10}{B} $\alpha \alpha p\Lambda$ cluster-model 
calculation \cite{Hiyama12} that results in 80~keV g.s. doublet excitation 
using CS $\Lambda N$ interactions. However, to do a good job on the level of 
$\sim$20~keV, one needs to include $\alpha$-breakup $\Lambda N$ contributions 
which are missing in this calculation.}}  
\end{itemize} 

It might appear unnatural that $\Delta E^{\rm CSB}_{\Lambda\Sigma}$ is 
calculated to be a sizable fraction of $\Delta E^{\rm CS}_{\Lambda\Sigma}$ 
in \lamb{8}{Li}, or even exceed it in \lam{10}{B}. This may be understood 
noting that the evaluation of $\Delta E^{\rm CSB}_{\Lambda\Sigma}$ involves 
a CSB small parameter of $\sim$0.03, see Eq.~(\ref{eq:OME}), whereas 
the evaluation of $\Delta E^{\rm CS}_{\Lambda\Sigma}$ involves a small 
parameter of $\sqrt{P_{\Sigma}}$ which is less than 0.05 for \lamb{8}{Li} 
and less than 0.025 for \lam{10}{B} in our $(\Lambda\Sigma)_{\rm e}$ coupling 
model, see Table~\ref{tab:pshell}.

\section{Conclusion} 

It was shown in this work how a relatively large CSB contribution of order 
250~keV arises in ($\Lambda\Sigma$) coupling models based on Akaishi's 
central-interaction $G$-matrix calculations in $s$-shell hypernuclei 
\cite{Akaishi00,Akaishi02}, coming close to the binding energy difference 
$B_{\Lambda}$(\lamb{4}{He})$-$$B_{\Lambda}$(\lamb{4}{H})~$=350\pm 60$~keV 
deduced from emulsion studies \cite{Davis05}. It was also argued 
that the reason for most of the $YNNN$ coupled-channel calculations done 
so far to come out considerably behind, with 100~keV at most by using 
NSC97$_{\rm f}$, is that their $\Lambda\Sigma$ channel coupling is 
dominated by strong tensor interaction terms. In this sense, the 
CSB-dominated large value of $\Delta B_{\Lambda}^4(0^+_{\rm g.s.})$ 
places a powerful constraint on the strong-interaction $YN$ dynamics.  

In spite of the schematic nature of the present ($\Lambda\Sigma$) coupling 
model of the $A=4$ hypernuclei, which undoubtedly does not match the high 
standards of solving coupled-channel four-body problems, this model has the 
invaluable advantage of enabling a fairly simple application to heavier 
hypernuclei, where it was shown to reproduce successfully the main CSB 
features as disclosed from the several measured binding energy differences 
in $p$-shell mirror hypernuclei. More quantitative work, particularly for 
the \lam{12}{C}--\lam{12}{B} mirror hypernuclei, has to be done in order 
to confirm the trends established here in the beginning of the $p$ shell 
upon relying exclusively on data reported from emulsion studies.  
Although the required calculations are rather straightforward, a major 
obstacle in reaching unambiguous conclusions is the unavailability of 
alternative comprehensive and accurate measurements of g.s. binding energies 
in mirror hypernuclei that may replace the existing old emulsion data.

\section*{Acknowledgements}  
Stimulating and useful exchanges with Patrick Achenbach, Ben Gibson, 
Johann Haidenbauer, Emiko Hiyama, Ruprecht Machleidt, John Millener, 
Andreas Nogga, Thomas Rijken and Hirokazu Tamura are gratefully 
acknowledged as well as financial support by the EU initiative FP7, 
HadronPhysics3, under the SPHERE and LEANNIS cooperation programs. 

\section*{Note added after publication} 
A recent LQCD work \cite{QCD15} evaluates the $\Sigma^0$--$\Lambda$ mixing 
matrix element $\langle\Sigma^0|\delta M|\Lambda\rangle$ to be smaller by 
about factor of 2 than the one used in the present work upon substituting 
PDG \cite{PDG14} baryon mass values in the Dalitz-Von Hippel \cite{DvH64} 
SU(3)$_{\rm f}$-symmetry expression (\ref{eq:deltaM}) which we copy here, 
\begin{equation}
\langle\Sigma^0|\delta M|\Lambda\rangle=\frac{1}{\sqrt 3}
(M_{\Sigma^0}-M_{\Sigma^+}+M_p-M_n). 
\label{eq:LQCD}
\end{equation} 
The resolution of this apparent discrepancy is that this much smaller 
value of $\langle\Sigma^0|\delta M|\Lambda\rangle$ in Ref.~\cite{QCD15} 
is obtainable simply by substituting LQCD-calculated mass differences 
in Eq.~(\ref{eq:LQCD}), particularly $M_n-M_p=2.70$~MeV which is about 
twice larger than the PDG value 1.29~MeV and which comes with a negative 
sign in (\ref{eq:LQCD}). We conclude that this LQCD determination of 
$\Sigma^0$--$\Lambda$ mixing is no better than their determination of 
the neutron-proton mass difference.

\end{document}